# Nanofabrication of spin-transfer torque devices by a PMMA mask one step process: GMR versus single layer devices

Anne Parge, Tore Niermann, Michael Seibt and Markus Münzenberg

IV. Physikalisches Institut, Universität Göttingen

We present a method to prepare magnetic spin torque devices of low specific resistance in a one step lithography process. The quality of the pillar devices is demonstrated for a standard magnetic double layer device. For single layer devices, we found hysteretic switching and a more complex dynamical excitation pattern in higher fields. A simple model to explain the resistance spikes is presented.

## I. Introduction

Starting with pioneering experiments in the 60ties, concepts were developed to treat the charge and spin degrees of freedom in transport experiments separately [1,2]. The spin-flip processes have also been identified in spin-diffusion experiments in the 80ties by Johnson and Silsbee [2], pioneering the research field of spin currents and spin-torque switching. The dominating spin-flip processes in metals originate from band mixing due to the spin-orbit coupling $\lambda \mathbf{SL}$. The mixing of the spin-up and spin-down band is connected with a spin flip probability which destroys the spin accumulation [4-6]. Nevertheless, currents injected from a ferromagnet into a normal metal show a high spin polarization which decays over an experimentally controllable range in the normal metal. Designing devices which adopt this kind of current in spinelectronic applications has become a challenging research field. First predictions on how a spin torque, which emerges with spin accumulation effects, can be used for the manipulation of the spin orientation have been made by Slonczewski [7] and Berger [8] in the 90ties. At that time nanofabrication techniques did not allow to produce devices with 100 nm diameter. Upon the development of nanofabrication techniques, spin transfer torque devices have been demonstrated to function in different geometries (mechanical point contact, lithographic point contact and lithographically etched nanopillar [9, 10]). The first demonstrations opened up a wide research field in magnetism in the past years. This led to new important developments: in the information technology industry, the new technique is implemented as a new switching mechanism for magnetic random access memories [11], whereas in high frequency research the devices are studied due to their potential application as nanometer microwave generators [12-14]. Even time domain measurements of the precession of the nanomagnet have been shown to be possible [15, 16]. In the following, we will demonstrate an alternative approach for the preparation of nanopillar structures in a one step process. In contrast to earlier work, we focus on the possibilities to characterize the pillars structurally in the first part. The preparation of lamella structures for transmission electron microscopy (TEM) by focused ion beam (FIB) makes a target preparation of a single pillar structure possible and gives insight into the crystalline growth as well as the element distribution. In the second part, spin current induced effects are studied in the standard Co/ Cu/ Co system in the current perpendicular to plane geometry as well as in a Co single layer device in order to demonstrate the functionality of the pillar devices prepared by the one step method. The single Co device does not possess a polarizing layer. In this case the Co acts as a polarizer itself and opens up a whole field of nonlinear physics. This allows to study the self-amplification of magnetic modes, whose fingerprint, a broad excitation spectrum, is observed in the resistance spectra here.

**II. Experimental method**

In the following we present a preparation process for magnetic pillar structures with diameters from 50 to 150 nm. They are fabricated in a simplified one step process. The lithographically defined hole and the e-beam resist itself, in our case Polymethylmethacrylat (PMMA), are used as a definition for the nanocontact preparation. We chose PMMA, because it is known to be a good electric insulator and a smooth buffer layer for the sequential film growth on top. Because of this properties, it is used in hard disk read/ write heads as an electrical spacer layer to insulate the coil section from the top and bottom yoke. Our technique is similar to the under-etched stencil masks used by Sun [17] with the advantage that the PMMA definition can be prepared by a one step process. A PMMA definition has been successfully used to structure a point contact on top of a multilayer stack by Rippard et. al. [13]. In their case the magnetic stack is an extended film and the current flow is defined by the nanocontact. Here both techniques are combined; the PMMA mask is not only used as a definition for the point contact, but also as an evaporation mask for the nanopillar itself. The drawback of nanostructures prepared by a definition mask technique is that at the edges of the pillar the film may not be well-defined or shorted due to shadowing effects at the side walls. On the other hand, pillar structures produced by $Ar^+$ etching may have the disadvantage that the side walls of the 100 nm diameter pillar structures can be object to defects arising from the etching process in some cases. The process of choice has to be optimized in both cases to avoid these effects. Further we used layer stacks as simple as possible (Co/ Cu/ Co/ Cu/ Pd and Cu/ Co/ Cu/ Pd). In this way we tried to avoid additional interface resistances. Additional layers are commonly used to stabilize the direction of the spin polarizing layer by inserting a natural or artificial antiferromagnet. Transmission electron microscopy (TEM) and scanning TEM (STEM) analyses have been carried out on a Philips CM200-FEG-UT operated at 200kV and equipped with a Si:Li detector (Link ISIS) for energy-dispersive X-ray spectrometry (EDXS).

**III. Experimental results**

   **A. Structure**

The pillar size and quality of the definition is controlled by cross sectional Scanning Electron Microscopy (SEM) on a cleaved substrate. A cross sectional view of one structure and the corresponding schematic layout are presented in Fig. 1. It shows the pillar structure from the side in between the substrate and the top Cu contact. The dimensions of the pillar diameter range from 150 to below 100 nm. Due to the cleavage process, the PMMA definition film and the Cu top contact are elastically deformed. To avoid shadowing effects by the side walls, the PMMA definition is rather steep as it can be clearly seen in the shape of the pillar as shown in Fig. 1. Secondly, the distance between the e-beam evaporation source and the sample holder amounts to 0.7 m, which guarantees a perpendicular angle of incidence of the impinging atoms. The angular opening is below $8 \times 10^{-6}$ degrees at the pillar position and thus the displacement towards the walls at the sides due to the divergence of the particle beam is theoretically below one Angstrom for an optimum alignment of the nanopillar template to the material evaporation source. The subsequent preparation steps are as follows: the bottom electrodes are prepared on a SiOx-substrate by mechanical masks. For optimum adhesion and low oxide formation a Ti/ Au bilayer is deposited. This is followed by spinning a 100 nm thick PMMA film onto the complete sample. On the middle of the bottom contact 100 nm spots are written by e-beam lithography for each pillar (Leo Supra 35, dot dose 0.075 pAs, beam voltage 20 kV). After the development step, the magnetic layer system is evaporated. The GMR stack is grown in a UHV chamber at a base pressure below $5 \times 10^{-10}$ mbar. Eight e-beam sources are available to have the flexibility to exchange the material combinations. In the last step, the pillar is filled by a thick 300 nm Cu film as a top contact using a cross stripe shadow mask to define the upper strip lines. Thus the critical lithography steps are essentially broken down into one write and one development step. All other steps for the definition of the electrical connection lines are accomplished by using simple shadow mask process steps. Nevertheless we are currently implementing a shadow mask system with the ability to increase the number of pillar contacts per substrate from five to thirty. In this way it will be possible to vary the thickness by using a wedge shaped thickness variation on one sample for exactly equal growth conditions.

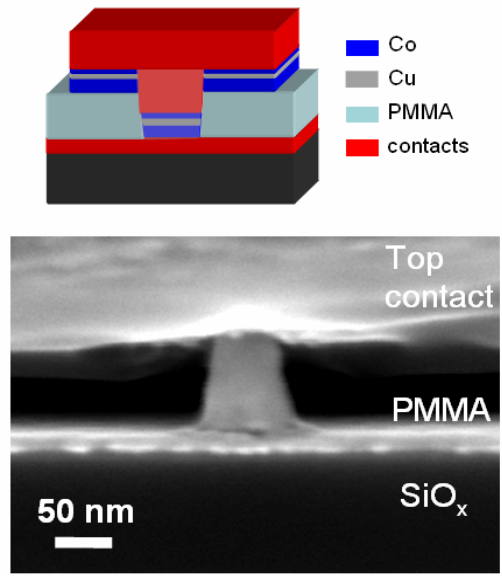

Fig. 1. Schematic layout of the pillar structure (top) and the SEM image of a cleaved substrate (bottom) show the pillar structure from the side in between the substrate and the top Cu contact. The dimensions of the pillar diameter range from 150 to below 100 nm. Due to the cleavage process, the PMMA definition film and the Cu top contact are elastically deformed.

The cross sectional SEM image of a cleaved substrate permits to study the quality of the pillar preparation and to optimize the e-beam lithography parameters. More complex is the target preparation of a single pillar structure by means of focused ion beam. The advantage is that this technique allows to have a look into the cross section of a pillar. Thereby the crystalline structure, the crystal orientation as well as the interface quality can be studied with atomic resolution. An image of the TEM analysis is shown in Fig. 2. The bright and dark field images are presented in a). In between the bottom Ti/ Au contact and the top Cu contact, a conic section of the pillar can be seen. It arises from the wedge-like cut by the ion beam used to prepare the TEM lamella. The concise geometry is drawn schematically in b). The PMMA, which shows only a weak contrast compared to vacuum, can be identified around the pillar structure. It is eroded by the $Ga^+$ beam very easily and forms characteristic holes observed to the left and right side of the pillar itself. Because of the background of the amorphous PMMA, the atomic structure of the pillar cannot be resolved in this series of images. Nonetheless they provide a profound amount of information on the structure. The diffraction contrast in the dark-field STEM-image is due to strain fields, small crystallites of different orientations and planar faults like twin boundaries (Fig. 2a). The crystalline growth itself seems to be coherent over the interfaces of the layer stack, since the contrast in bright and dark field images does not change from one layer to the other. The layer stack can be identified from X-ray maps obtained in the STEM mode of the electron microscope with a probe size of about 1nm. (Fig. 2c). The TEM images reveal that the GMR layer stack of the pillar structure is well-defined and no shadowing effects are observed. Due to the conical shape of the cross section through the pillar, the image contrast is getting weaker for the layers in the upper part of the pillar and finally fades out. It would be desirable to distinguish between hcp and fcc Co, since anisotropies are strongly related to the crystalline lattice as well as lattice distortions. Future improvements of the target preparation technique will allow new and unique characterization possibilities, whose results are of utmost significance to micromagnetic modeling. Information on the growth or structure after the lithographic processing has not been accessible for the major part of publications in that field yet, especially not on an atomically resolved level. This will be possible by improving the target preparation process outlined here.

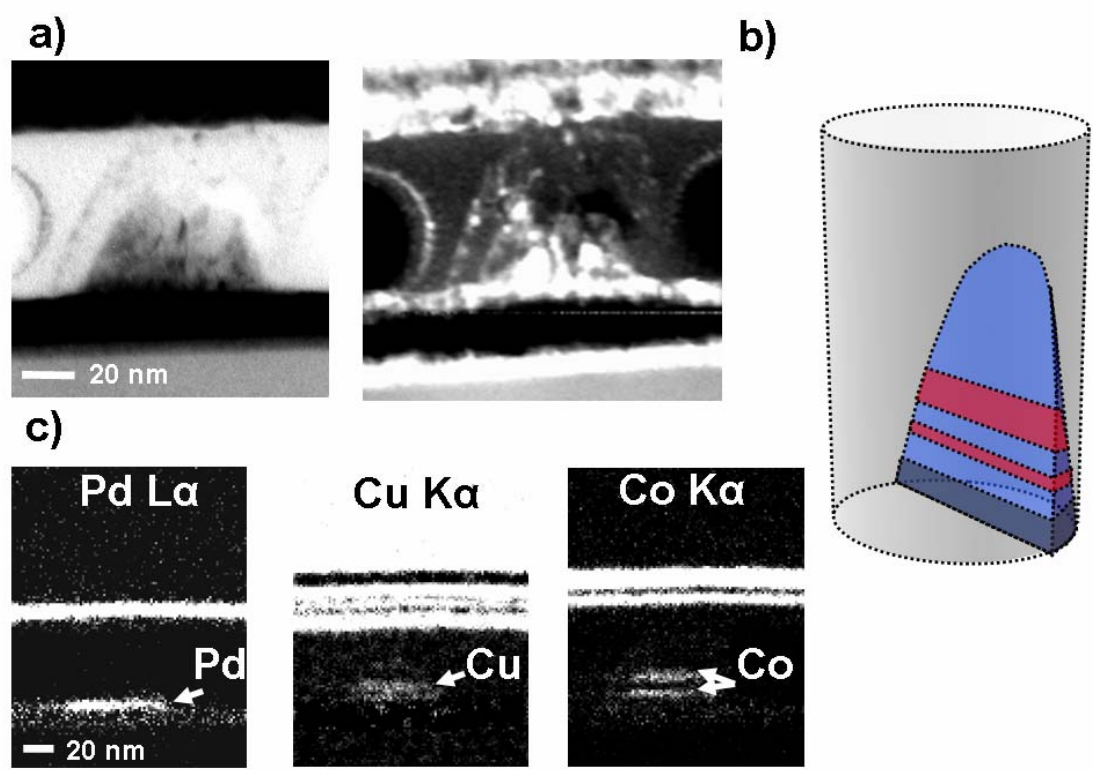

Fig. 2. TEM images of a trilayer Co/ Cu/ Co pillar with a Pd spin diffusor. In a) the STEM bright and dark field images are shown. The cross section was prepared by focused ion beam. A corresponding schematic drawing is shown in b). The wedge-like shape of the TEM lamella results in a section through the entire pillar under a constant angle. The conic section is seen in the images. In c) the EDXS maps obtained from the Pd $L_\alpha$, Cu $K_\alpha$ and Co $K_\alpha$ radiation are shown; the layer stack can be identified within the pillar and on top of the PMMA.

## B. Trilayer structures

At first, the transport properties of the Co/ Cu/ Co trilayer system will be demonstrated in the high-field limit with the field applied out-of-plane along the pillar direction. Within the experimental setup, a field range of up to 9 T and temperatures from 10 K up to 400 K can be investigated. The field can be rotated from the in-plane to the out-of-plane direction to fully characterize the magnetic properties. Because of the low resistance of the elements, a dual lock in technique is used to measure dV and dI simultaneously in the four point geometry. The same kind of setup was developed and has already been used successfully in Ref. 18.

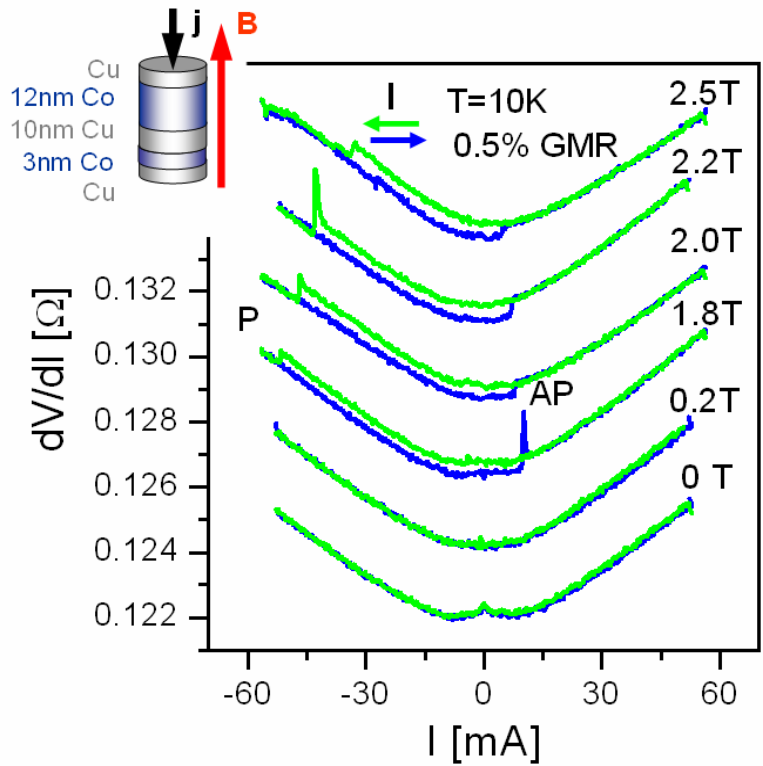

Fig. 3. Current-induced switching of a 12nm Co/ 10nm Cu/ 3nm Co trilayer pillar with a Pd spin diffusor. The spectra are shifted vertically by 2 m$\Omega$ for different fields. The magnetic field is applied parallel to the pillar (out-of-plane).

For a 12nm Co/ 10nm Cu/ 3nm Co sample we restrict ourselves to the current-induced switching effect in the high field range up to 3 T, which has already been investigated in Ref. 19 and 20. In zero field, the ground state for a 12 nm thick Co polarizing layer is a magnetic vortex configuration. The 3 nm thick Co layer is almost homogeneously magnetized in plane. Hysteretic switching sets in for fields above a certain critical value which corresponds to the demagnetization field of about $4\pi M_s N_z \approx 1.5 \text{ T}$. Such a correlation with the demagnetizing field has also been observed in previous works: in order to have a spin polarized current with a net polarization, the field has to be large enough to pull the magnetization out of the layer plane. As a second observation, the magnetoresistance effect increases slightly with larger external fields between 1.8 and 2.2 T, since the magnitude of the GMR effect is determined by the angle between the directions of magnetization in the two layers. The resistance change for the structure shown is very sharp. From the parallel to the anti-parallel configuration the change amounts to 0.5%. For positive currents (electron current flow through the thin layer to the polarizer), the configuration switches from the parallel (low resistance, P) to the anti-parallel state (high resistance, AP) at a critical current $j_c A$ of a few mA only ( $j_c$ critical current density, $A$ area). Coming from the AP state, a much higher $j_c$ is needed to switch to the P configuration. Along with the switching an intermediate state emerges, which can be identified by a large spike in the differential resistance. It directly indicates a spin wave instability marking the onset of the switching. This additional feature in dV/dI, which does not arise from the GMR for parallel and antiparallel magnetization orientations in the layers, will be discussed further in connection with the magnetoresistance effects observed in the single Co layer device. With increasing applied field, the critical current density $j_c$ decreases as it is displayed in Fig. 3. The value of $j_c$ is determined by the spin torque required to switch from the parallel to the antiparallel configuration, thus to overcome the applied fields, the dipolar fields and the crystalline anisotropy. The hysteresis is shifted from zero which is expected due to the asymmetry introduced by the thick Co layer acting as a spin polarizer. In contrast to Ref. 19, a much wider hysteresis is observed. The anti-parallel orientation is also stable for zero applied currents.

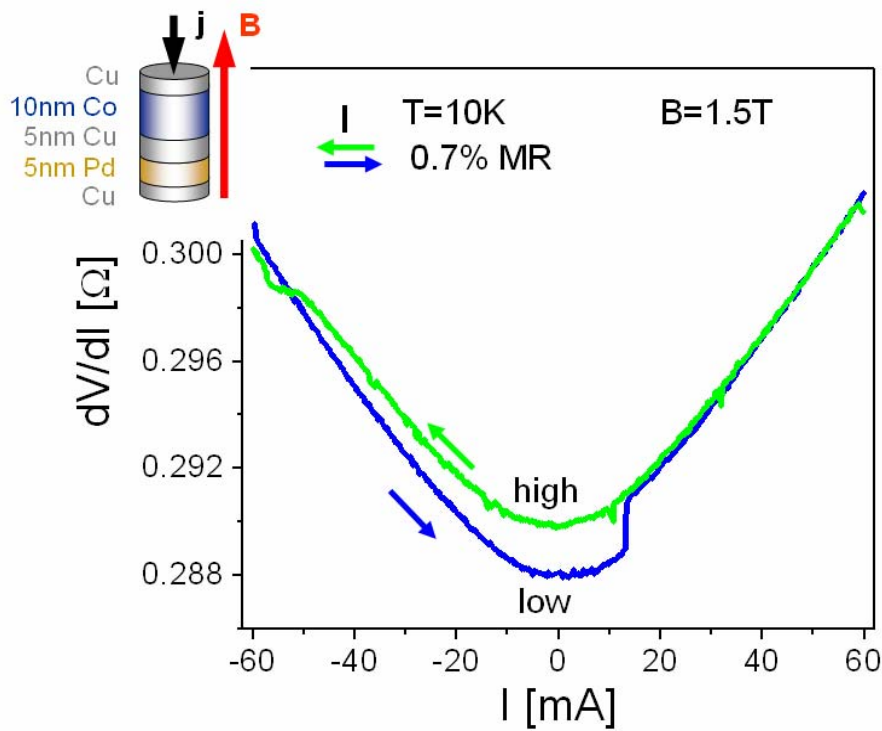

Fig. 4. Current-induced switching for a 10 nm single Co layer with a Pd spin diffusor.

## C. Single layer structures

In addition, we present transport measurements for a single 10 nm Co layer system. The build-up of magnetization dynamics and spin-wave modes in a single layer has been first observed and modeled by Stiles [21] and Polianski [22]. Here the magnetic layer itself acts as the polarizing layer. For a field value of 1.5 T hysteretic switching as shown in Fig. 4 is observed. Furthermore, the differential resistance versus current curve also shows a characteristic asymmetry. It is shifted from zero which is expected due to the asymmetry introduced by the Pd diffusor. Another characteristic trait is the abrupt resistance change of the hysteresis for positive current bias (electrons passing through the spin diffusor before entering the Co layer), as opposed to the smooth transition for negative currents (electrons approaching the Co layer from the pure Cu leads). In the first case the spin accumulation is built up by the reflected electrons at the upper interface of the Co layer. In the second case the Co layer acts as a polarizer itself due to the difference in conductance for both spin channels in the Co for the transmitted current. Details on the origin of the magnetoresistance effects in a single layer have already been discussed by Özyilmaz [23]. For a homogenous magnetization, one spin channel is reflected at the interface, thus blocked, and as a result a spin accumulation is built up at the interface. This spin accumulation can be reduced by spin diffusion along the interface if there is a transverse inhomogeneity in the magnetization. The resistance change can be understood by a resistor model assuming two resistances for both spin channels in parallel. Then a *decrease* in resistance corresponds to the opening up of a lower resistance channel in parallel. This is the case if two opposite magnetization directions are present in the Co film (out of plane component parallel to the B field and antiparallel to B). However, an *increase* in resistance within this model then corresponds to an additional resistance in series. In this case a domain wall like resistance originating from the gradual rotation of the magnetization along the current direction is existent. The gradual rotation must be within the spin relaxation length in order to cause a significant change in the resistance (adiabatic limit) [24]. Two models to relate a magnetization configuration to the resistance change are described below, although a thorough description does not seem to be possible.

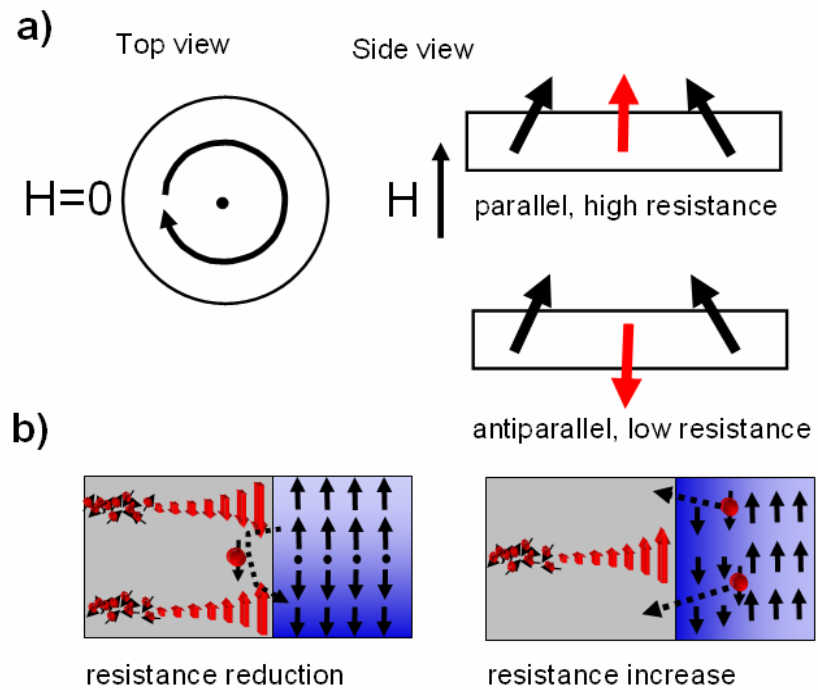

Fig. 5. a) Schematic drawings for the magnetization configuration with different external fields and current bias. b) Model for the resistance reduction by spin diffusion along the interface, transverse to the current direction (left) and the resistance increase by an inhomogeneous magnetization along the current direction (right).

The simple model to explain the resistance change is depicted in Fig. 5, based on ideas first developed by Özyilmaz et. al. [23]. In Ref. 23, the elements were elliptical, so that the authors assumed a configuration with three vortices for this structure. Here the model is adapted to the circular geometry. In the ground state, the Co film in the pillar is in a vortex configuration. By applying a field perpendicular to the film plane, pointing along the direction of the external field, the vortex core is dilated with increasing out-of-plane field. Within this model, it is assumed that the outer part of the vortex depicted by the black arrows is fixed, while the inner part marked by the red arrows is supposed to be flexible. The magnetization is tilted out-of-plane until the demagnetization field is reached. With increasing current, a spin torque acts on the flexible inner part of the vortex and eventually switches the magnetization into an antiparallel configuration of lower resistance. Coming from the high resistance state, the resistance is reduced at a critical current of approximately -48 mA and the mode with the core of the magnetization is flipped into the antiparallel configuration. Upon lowering the current, it is remarkable that the disk stays in the metastable low resistance state even for zero current, probably stabilized by the reduced dipolar field in this configuration. For a positive current of 17 mA, the inner part of the vortex is flipped back to the all parallel high resistance state again. It is obvious that a macrospin model will not work for the single layer case. While the vector model for two distinct regions (outer part and inner part of the vortex) as described above is only the simplest way to address the problem, a far more elaborated manner to study the accessible modes in such a structure are micromagnetic simulations. These have been performed in Ref. 25: Here the mode spectrum has been studied in terms of current-induced excitations for a 15 nm thick Co disk magnetized out-of-plane, excited in the simulation by a strongly inhomogeneous field pulse. Two distinctly different modes have been explored. In addition to the lateral magnetic inhomogeneity (transverse to the direction of current flow), the mode spectrum has been classified by an even and an odd mode in z direction (longitudinal to the current flow). While the first mode will result only in a resistance reduction, the latter will lead to a resistance increase because of its longitudinal inhomogeneity in the magnetization. The second odd mode was proposed to have the lowest excitation threshold, since the asymmetry in z (in addition to the asymmetry introduced by the Pd diffusor) lowers the threshold current.

An alternative possibility to describe the observed hysteretic effects is to assume a hysteretic switching between a vortex and a vortex-free state of the magnetization (c or s state). In this case the vortex state lowers the resistance due to the transversal magnetic inhomogeneity; all magnetization directions are present and the lateral spin accumulation will be reduced. A support for this model comes from micromagnetic simulations. Circular Co structures with diameters of 80 and 100 nm and a varied thickness have been investigated using the OOMMF simulation package (cell size 1 nm) [26]. The transition thickness from the almost homogeneously magnetized thin Co films to the vortex state for the thicker films is found to be at 6.8 nm and 9.5 nm, respectively. For the 10 nm thick single layer device, the vortex structure has the lower energy (436 kJ/m$^3$) close to the homogeneously magnetized c or s state (465 kJ/m$^3$). Thus both are almost degenerated in energy and switching between them is facilitated. However, micromagnetic simulations including the spin accumulation effects are necessary to explore the origin of the observed hysteretic switching in greater detail in the future. Such codes are currently developed and thus investigations in this field have only just begun.

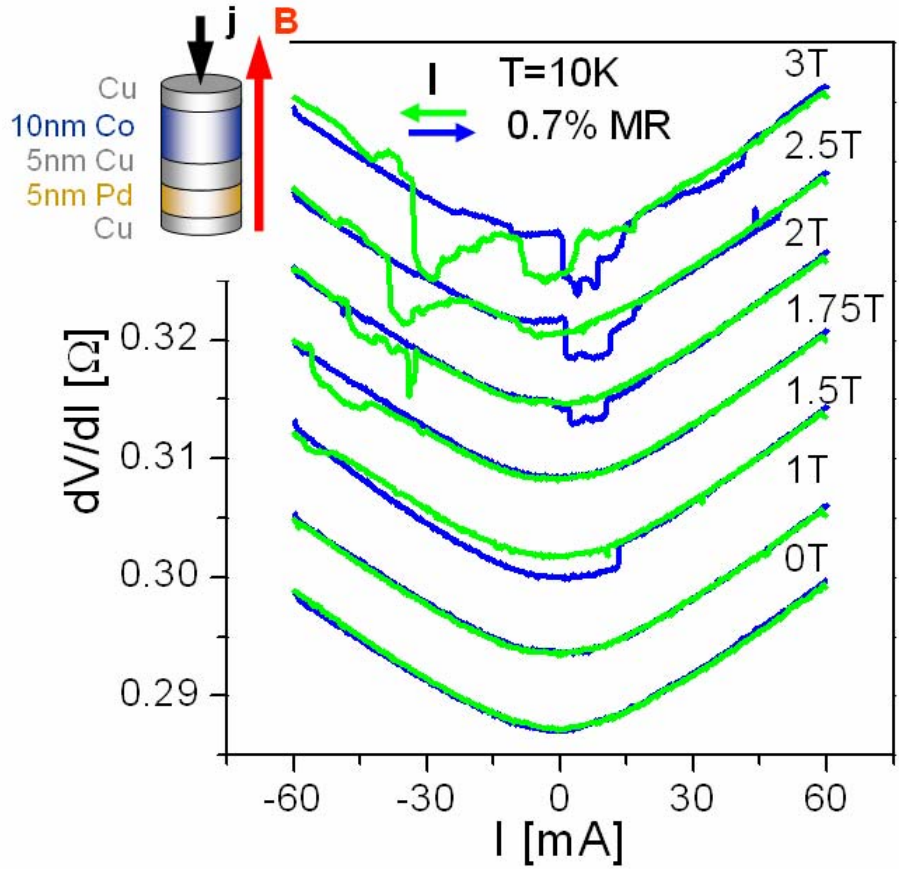

Fig. 6. dV/dI spectra versus current for various applied fields (T=10 K). The spectra are shifted vertically for clarity.

In Fig. 6 we show how the differential resistance spectra evolve in increasing applied magnetic fields. The magnetic field is again applied along the pillar direction. As expected, magnetization dynamics and magnetic switching originating from the spin torque which is exerted onto the magnetization is observed only for higher fields. In 0 to 1 T external field no excitations are found and, for fields larger than that, the critical currents decrease with the applied field. This is contrary to Ref. 23, where magnetic excitations are observed also for low fields and the critical current densities $j_c$ show a linear increase with the applied field. The linear dependency between $j_c$ and the mode frequency $\omega_i$ are quoted from [21]. Here, the critical current $j_c$ for the excitation of a mode with the frequency $\omega_i$ has to be increased in order to overcome the additional damping proportional to $\frac{\partial \omega_i}{\partial H}$. If the mode frequency $\omega_i$ is found to increase linearly with the field $H$, also $j_c$ is increased linearly. The model is strictly valid for the mode spectrum in the saturated case only (M out-of-plane) [21]. To explain the differences to our observation, again micromagnetic simulations can be consulted. In zero field, the polarizing layer is in a magnetic vortex configuration. As the field is increased, first the vortex core is considerably widened (the vortex width is 10 nm, dilating up to 20 nm) and the out-of-plane magnetization component increases gradually. The vortex vanishes for about 1.3 T. In order to have a spin polarized current with a sufficient net polarization, the external magnetic field has to be large enough. This also explains the emergence of magnetization dynamics for out-of-plane field values corresponding to the saturation field $H_s$, which is similar to the observations for the trilayer system: since no net in-plane spin accumulation is built up in the vortex configuration, it is reasonable to connect the onset of the magnetic excitations with the vanishing of the vortex core. For higher fields, the hysteretic switching evolves to broad resistance dips at the position of the critical current where the onset of the hysteric switching was before. Similar effects have been presented for the trilayer case in Ref. [27]. As a second difference to the excitations described by Özyilmaz et. al. [23] for the single layer case, the critical current is not increasing with the applied field. On the contrary, a reduction of the threshold current is observed. While the critical current stays the same for small positive values, the critical current at negative bias is reduced with increasing field. This indicates the existence of large angle excitations [23]. Upon larger deviations from the ground state the spin torque can act more efficiently on the magnetization and the critical currents are reduced. Most intriguing is the wide current range of the excitation, overlaid by various peaks forming a rather broad minimum. In a blow-up, small well-defined steps of almost equal size can be observed. These seem to correspond to the onset of an excited mode or to additional vortex cores moving into the disk. The more complex excitation patterns in the single layer case can be understood by comparing the excitation mechanisms. In the trilayer case, the polarizing layer is fixed, and thus it defines the spin current polarization. In the case of the single layer system, the Co layer itself defines the spin accumulation. When the switching or precession sets in, the spin accumulation is reduced at the same time. This results in nonlinear effects in the excitation pattern. Since the direction of the inhomogeneous magnetization along the metal/ ferromagnetic interface is coupled to the interfacial spin accumulation and vice versa an inhomogeneity can be identified. It should be possible to amplify various magnetic eigen modes of the Co film over a large range of currents. Complex and more chaotic excitation patterns have already been treated in theory [10, 27-30]. The identification of these modes is not possible in our experiments by now. Time resolved experiments with ps resolution projected in the future by introducing an ultrafast photoconductive switch, will make it possible to study these complex excitations in real time.

**IV Conclusions**

In summary, we presented a new straightforward technique to prepare nanometer sized pillar structures for the investigation of spin current-induced switching effects. The preparation technique is verified by an element-resolved cross sectional study of the pillar region and dV/dI spectroscopy experiments for the trilayer geometry for various field values in the field out-of-plane geometry. For the single layer a relatively wide range of magnetic excitations is identified forming a rather broad resistance minimum. We believe these originate from self-amplification effects of the eigen modes of the single Co film and open up an interesting research field in the future.

**Acknowledgements**

Support by the Deutsche Forschungsgemeinschaft within the priority program SPP 1133 is gratefully acknowledged. We thank Dr. P Wilbrandt and V. Radisch for the preparation of the TEM lamellas.